\input{aipcheck}

\documentclass[
  ,draft            
  ]
  {aipproc}

\layoutstyle{6x9}

\begin{document}

\author{S.A.~Bonometto, R. Mainini, L.P.L. Colombo}{
  address={Physics Department, Milano--Bicocca University,
Piazza della Scienza 3, 30126 Milano (Italy) \&},
  address={I.N.F.N, Sezione di Milano--Bicocca},
  email={bonometto@mib.infn.it},
}


\title{The Dark Side and its Nature}

\date{2006/8/21}


\begin{abstract}
Although the cosmic concordance cosmology is quite successfull in
fitting data, fine tuning and coincidence problems apparently weaken
it. We review several possibilities to ease its problems, by
considering various kinds of dynamical Dark Energy and possibly its
coupling to Dark Matter, trying to set observational limits on Dark
Energy state equation and coupling.
\end{abstract}

\maketitle

\it
\noindent
{1.~~Introduction}
\rm
\vglue .2truecm

\noindent
Until a decade ago two options were in competition: the world could be
either SCDM or 0CDM. The former cosmology had matter density and
deceleration parameters $\Omega_{om} \simeq 1$ and $q_o = 0.5$; the
latter had $\Omega_{om} \simeq 0.2$--0.3 and $q_o \sim 0\,
$. $\Omega_{om} \simeq 1$ was supported by COBE data and agreed with
generic inflationary predictions; $\Omega_{om} \simeq 0.3$ was
supported by evolutionary data and inflation made it acceptable
without horrible fine tunings. When data became too far from SCDM, the
mixed model variant
\cite{valda} became popular.

Then SNIa data \cite{riepea} required $q_o \sim
-0.6$--0.7. $\Omega_{om}$ could still be $ \simeq 0.3$ if the gap up
to $\Omega_{o}=1$ was covered by a {\it substance} with $p
\simeq -\rho$, the Dark Energy (DE). The $\Lambda$CDM models, 
considered until then little more than a smart counter--example, begun
their fast uprise to become the Cosmic Concordance Cosmology (CCC).
Then, soon after SNIa data, deep sample analysis \cite{deep} and fresh
CMB data \cite{cmb} converged in confirming that $\Omega_{om} \simeq
0.3$ with $\Omega_{o} \simeq 1\, $ and the CCC became a must.

Only a minority were however happy with $\Lambda$ being the
cosmological constant. Thus, CCC brought the problem of DE nature.
Most of this paper deals with DE being a self--interacting field,
either fully decoupled or decoupled from any other component apart
Dark Matter (DM): dynamical and coupled DE (dDE and cDE),
respectively.

\begin{figure}
\includegraphics*[height=2.5cm,width=7.5cm]{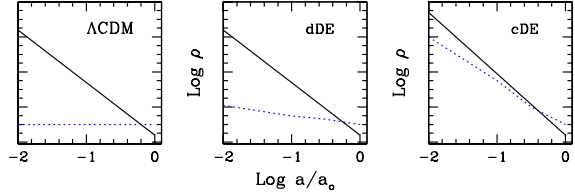}
\caption{DM and DE densities in $\Lambda$CDM, dDE amd cDE cosmologies.
The solid (dotted) line shows DM (DE) densities in the different
classes of models. }
\label{s3}
\end{figure}

Fig.~\ref{s3} illustrates why we favor these options. If DE pressure
and density meet the condition $p_{de} = -\rho_{de}\, $ exactly, then
$\rho_{de}$ is constant: backward in time, DE rapidly becomes
negligible and we wonder {\it why just today} it became relevant. This
is the {\it coincidence} problem. But, if DE is false vacuum, at the
end of the transition, $\rho_{de}$ should have became $\sim 1:10^{54}$
of its pre--transition value. This is a typical {\it fine tuning}
problem.

The latter problem vanishes in dDE models. In the interaction
potential no fine--tuned scale appears. However, as shown in panel
dDE, the coincidence problem is not eased. Then, if DE suitably
interacts with DM, densities can evolve as in panel cDE, and also the
coincidence problem is eased: DE and DM have had comparable densities
since long.

We became accustomed to cosmologies with various components of similar
density, {\it e.g.}, baryons and DM. CCC now requires also DE to fall
in the same range. This would not be an extra requirement if DE and DM
are just two different aspects of a {\it Dark Side}, their interaction
being a signature of a common nature. An example is the {\it dual
axion model} \cite{dual}, but here we approach the question from the
phaenomenological side.

We outline soon that DM--DE coupling predicts a significant baryon--DM
segregation in non--linear structures.  Hydrodynamics succeeds in
explaining most segregation effects observed in the real word, by
tuning suitable parameters. If the required tunings conflict with
other data or other observations require a different behavior of
baryons and DM still before the onset of hydro, the DM--DE coupling
would find an observational support.

Constraints on cDE from CMB data were first discussed by
\cite{amendola}. In Section 2 the results of a fit to WMAP1 data
\cite{colombo1} are reported. Discussing non--linear constraints is a
harder task. Even predictions based on the Press \& Schechter or Sheth
\& Tormen \cite{PSST} formalism are hard to obtain. In Section 3 we
shall explain why it is so. In Section 4 we shall then use ST
expressions to predict mass functions. Our conclusions are in Section
5.

\vskip .4truecm
\noindent
{\it 2.~~Background and linear fluctuations}
.
\vglue .2truecm
\noindent
If DE is a scalar field $\phi$, self--interacting through a potential
$V(\phi)$ \cite{wetterich}, \cite{RP}, it is
\begin{equation}
\rho_{de}=\rho_{k,de}+\rho_{p,de} \equiv {{\dot{\phi }}^{2}/2a^{2}}+V(\phi ),~
~~~
~~~
p_{de}=\rho_{k,de}-\rho_{p,de} \equiv w\, \rho_{de}~.
\end{equation}
Then, if dynamical equations yield $\rho _{k,de}/V\ll 1/2$, it is
$-1/3 \gg w>-1$. We use the background metric $ ds^{2} = a^{2}(\tau)
(-d\tau^{2}+ dx_i dx^i ) $; dots indicate differentiation with respect
to $\tau $ (conformal time). This DE is dubbed {\it dynamical} (dDE)
and much work has been done on it, also aiming at restricting the
range of acceptable $w(\tau)$'s, so gaining an observational insight
onto the physics responsible for the potential $V(\phi)$.
\begin{figure}
\includegraphics*[height=6cm,width=6cm]{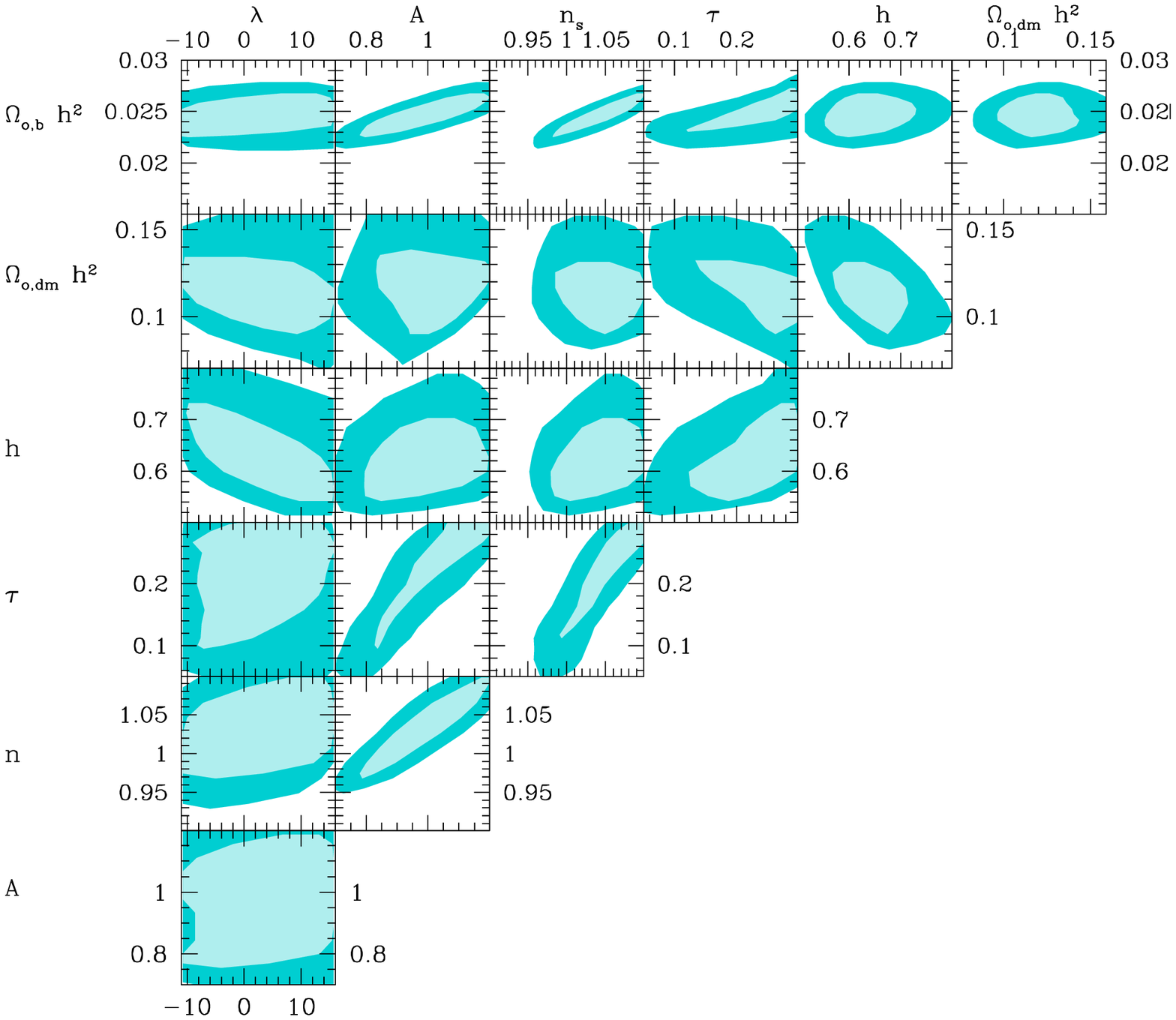}
\includegraphics*[height=6cm,width=6cm]{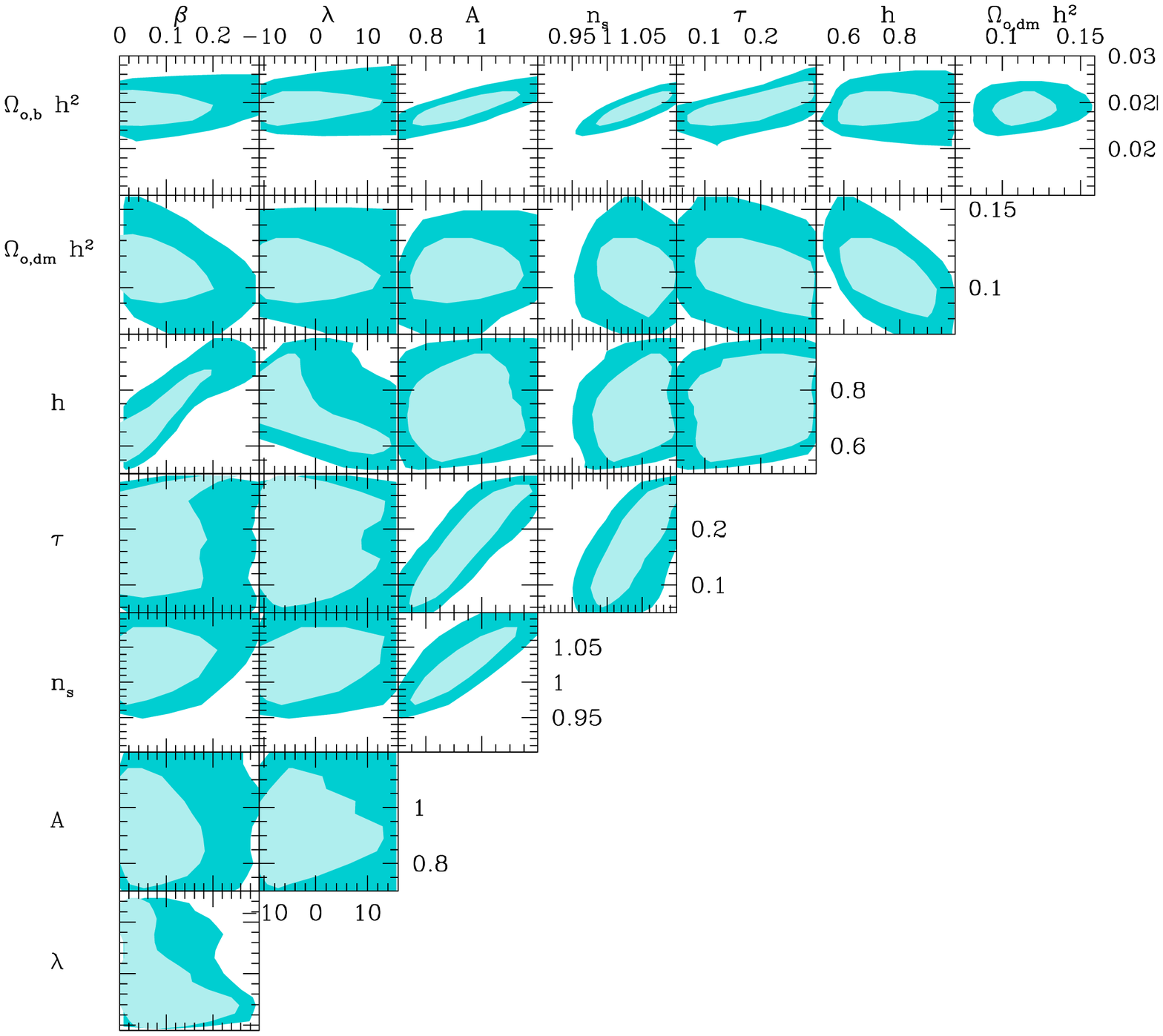}
\caption{1-- and 2--$\sigma$ limits on model parameters, obtained from 
a MCMC fit of dDE and cDE (SUGRA) cosmologies with WMAP1. Here
$\lambda =\log_{10}(\Lambda/{\rm GeV})$; $A$, $n_s$, $\tau$ are
fluctuation amplitude, primeval spectral index, opacity.  $\Lambda$
turns out to be substantially unconstrained. For dDE, $n_s$ is greater
than for $\Lambda$CDM, but all parameter values are compatible with
$\Lambda$CDM values within $\sim 2$ $\sigma$'s. For cDE, the coupling
$\beta$, at 2--$\sigma$, should be $< \sim 0.3$. }
\label{mcmcdec}
\end{figure}

In this paper we shall consider the potentials

\vglue .1truecm
$
SUGRA :~~
V(\phi) = (\Lambda^{\alpha+4}/\phi^\alpha) \exp(4 \pi \phi^2/m_p^2)
~,~~~~~
RP :~~
V(\phi) = \Lambda^{\alpha+4}/\phi^\alpha  ~
$

\vglue .1truecm
\noindent
(\cite{brax}, \cite{RP}), admitting tracker solutions and yielding two
opposite $w(\tau)$ behaviors: nearly constant for RP, fastly varying
for SUGRA. Then, the effects we find should not be related to the
shape of $V(\phi)$ but to the coupling. Most results are shown for
$\Lambda = 10^2\, $GeV; minor shifts occur when varying
$\log_{10}(\Lambda/{\rm GeV})$ in the 1--4 range.

DM--DE coupling, fixed by a suitable parameter $\beta$, modifies
background equations for DE and DM as well as those ruling density
perturbations (see, e.g.,
\cite{amendola}).  

Linear codes, modified by in this way allow predictions on CMB
anisotropies and polarization. In this way we fitted dDE and cDE
models to WMAP1 data
\cite{colombo1}. In Figs.~\ref{mcmcdec}
 MCMC limits on model parameters are shown, for the dDE and cDE with
SUGRA potential. The $\chi^2$ (likelihhod) is 1.066 (4.7$\, \%$) both
for $\Lambda$CDM and cDE, 1.064 (5$\, \%$) for dDE, with a marginal
improvement. More significantly, cDE is not worse than $\Lambda$CDM.

\vskip .4truecm
\noindent
{\it 3.~~Non--linear Newtonian approximation}
\vglue .2truecm

\noindent
While CMB analysis is based on linearized eqs., if we
deal with scales well below the horizon and non--relativistic
particles, the restriction $\delta_{c,b} \ll 1$ can be lifted. An
alternative {\it Newtonian} approximation is then licit, within which
tensor gravity and scalar field cause overlapping effects adequately
described by assuming:

\noindent
(i) DM particle masses to vary, so that
$
M_{c} (\tau) = M_{c} (\tau_i) \exp[-C (\phi - \phi_i)].
$

\noindent
(ii) Gravity between DM particles to be set by $G^* = \gamma G$.

\noindent
Here
$
C = \sqrt{16\pi G/3} ~\beta ~,~~
\gamma = 1+4\beta^2/3 
$ (see, e.g., \cite{maccio}). This approach allowed us to study
spherical {\it top--hat} fluctuations \cite{mainini6}, so predicting
the cluster mass functions in cDE models \cite{maibon6}, and to
perform n--body simulations \cite{maccio}.

The main novel feature of cDE cosmologies we outline in this way is
DM--baryon segregation, a strong effect, visible in the evolution of a
spherical {top--hat} fluctuation. 

Although in all cosmologies, apart SCDM, the spherical growth must be
studied numerically, in cDE, just because of the ongoing segregation,
the numerical approach is essential. In all cases apart cDE, the only
variable describing a {\it top--hat} is its radius $R$, set to expand
initially as scale factor $a$. Then, the greater density inside the
top hat slows down $R$ in respect to $a$, so that the decreasing
$\rho(<R)$ increases in respect to the average $\rho$.  At a time
$t_{ta}$, when $\Delta = \rho(<R)/\rho$ attains a suitable value
$\chi$, $R$ starts decreasing, to formally vanish within a finite time
$t_c$. For sCDM, $\chi = (3\pi/4)^2$ and $t_c/t_{ta} = 2$. In other
non--cDE models, $\chi$ and $t_c/t_{ta} $ take slightly different
values.

However, when $\rho (<R)$ increases, unless the heat produced by the
$p\, dV$ work is radiated, virial equilibrium is soon attained, and
any realistic fluctuation stops contracting at a radius $R_v$. In
sCDM, $R_v/R_{ta} = 1/2$ and, taking into account the symultaneous
growth of $a$, the virial density contrast is then $\Delta_v = 32\chi
\simeq 180$, indipendently of $t_c$. Values are slightly different
for other $\Lambda$CDM and dDE and a usual assumption is that the 
system relaxes into its virialized configuration just within $t_c$.

$\Delta_v$ values for $\Lambda$CDM and dDE were obtained by
\cite{Lahav_brian} and \cite{io,collapse1}, respectively. The same 
problem was treated in \cite{mainini6} for cDE. In this case, $R_b$
and $R_c$ (baryon and DM radii) initially grow as the scale factor,
but $R_{b}$ soon exceeds $R_{c}$, because of the stronger effective
gravity; peripheral baryons then leak out from $R_c$, so that DM no
longer feels their gravity, while baryons above $R_c$ feel the gravity
also of {\it external} DM layers. Then, above $R_c$, the baryon
profile is no longer {\it top--hat}, while a fresh perturbation in DM
arises. Re--contraction will then start at different times for
different components and layers. Similarly, virialization conditions
are fulfilled earlier by inner layers, although outer layer fall--out
shall later perturb them so that the onset of virial equilibrium is a
multi--stage process.  Furthermore, when the external baryons
fall--out onto the virialized core, richer of DM, they are accompanied
by DM materials originally outside the top--hat, perturbed by baryon
over--expansion.

Each layer and substance, in such system, feels a different force;
each shell needs then to be considered separately; this is why 
the numerical problem is far more intricate.
\begin{figure}
\includegraphics*[width=6cm]{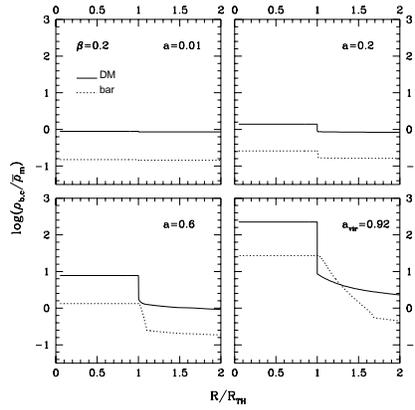}
\caption{Density profiles at different $a$ values for $\beta = 0.2$
model. Notice the progressive deformation of the baryon profile
(dotted lines) in respect of the DM profile (solid line).}
\label{profili02}
\end{figure}

The collapse stops at virialization. Fig.~\ref{profili02} assumes the
growth to stop when all DM originally in the top--hat, and baryons
inside it, virialize. Already at $a = 0.2$, before the turn--around,
the baryon top--hat boundary is no longer vertical. At $a=0.6$
($\sim$turn--around), a similar effect is visible also for DM. The
effect is even more pronounced at $a_{vir } \simeq 0.92$.  The same
effects, just slightly weeker, are present also for $\beta = 0.05$

Fig.~\ref{profili02} shows also that baryons leak out from the DM
top--hat. For $a \sim 0.92$, when DM and inner baryons virialize,
$\sim 10\, \%$ ( $\sim 40\, \%$) of baryons are out of the top--hat,
if $\beta = 0.05$ ($\beta = 0.2\, $). These values increase by an
additional $ 10\, \%$ in the RP case.

The virialization of materials within $R$ requires that $ 2\, T(<R) =
R\, dU(<R)/dR\, $. Kinetic and potential energies have fairly
straightforward expressions. The only point to outline is that DM--DE
energy exchanges, for the background, are accounted for by background
evolution eqs., so that, when fluctuations are considered, the
background contribution must not considered again.

\vskip .4truecm
\noindent
{\it 4.~~Mass functions in cDE theories}

\vglue .2truecm
\noindent
Let us compare the actual growth of a top--hat fluctuation with its
growth if we assume linear equations to hold, indipendently of its
amplitude. While the {\it real} fluctuation abandons the linear
regime, turns--around, recontracts and recollapses at the time
$\tau_{rc}$, {\it linear} equations let that fluctuation steadily
grow, up to an amplitude $\delta_{rc}$ at the time $\tau_{rc}$.

The linear evolution does not affect amplitude distributions. If they
are initially distributed in a Gaussian way, at $\tau_{rc}$ we can
still integrate the Gaussian from $ \delta_{rc}$, so finding the
probability that an object forms and virializes. This is the basic
pattern of the PS--ST approach, that we apply also to cDE cosmologies.
For them, however, $\tau_{rc}\, $ is different for DM and
baryons. {\it Viceversa}, if we require $\tau_o$ to coincide with
$\tau_{rc}$, we have two different {\it linear} amplitudes,
$\delta_{rc}^{(b)}$ and $\delta_{rc}^{(c)}$, yielding recollapse at
$\tau_o$ for all baryons or all DM.

The $\beta$ dependence of $\delta_{rc}^{(c,b)}$ at $z=0$ and their
$z$ dependence are given in \cite{maibon6}.  Setting $\nu_M =
\delta_M/\sigma_M$, the PS differential mass function then reads
\begin{equation}
\psi (M) = \sqrt{2 / \pi} (\rho/ M)
\int_{\delta_{rc}/\sigma_M}^\infty d\nu_M
~(d\nu_M / dM) \, \nu_M \exp\{-\nu_M^2 /2 \} ~,
\label{ps}
\end{equation}
using here $\delta_{rc}^{(c)}$, $\delta_{rc}^{(b)}$, or any
intermediate value, according to the observable to be fitted.  The ST
expression is obtainable from eq.~(\ref{ps}) through the replacement

\noindent
$
\nu_M \exp (-{\nu_M^2 \over 2}) \to
{\cal N}'\, {\nu'}_M (1+{\nu'}_M^{-3/5})\, \exp (-{{\nu'}_M^2 \over 2})$,
~with
$
{\cal N}'=0.322,~{\nu'}^2_M = 0.707\, \nu^2_M,
$

\noindent
meant to take into account the effects of non--sphericity
in the halo growth.

In $\Lambda$CDM or dDE, the mass $M$ in ST expressions {\it is} the
mass originally in the top--hat. In cDE, indipendently of the value
taken for $\delta_{cr}$, {\it virialized systems will be baryon
depleted}.  In fact, mass function built by taking $\delta_{cr}^{c}$,
concern objects before a part of the initial baryon content has fallen
out. But, even if we wait for a total or partial baryon fall out, by
using $\delta_{cr}^{b}$ or an intermediate value, the initial baryons
come back carrying with them DM layers initially external to the
top--hat. A prediction of cDE theories, therefore, is that
$\Omega_c/\Omega_b$, measured in any virialized structure, exceeds the
background ratio.

\begin{figure}
\includegraphics*[width=6cm]{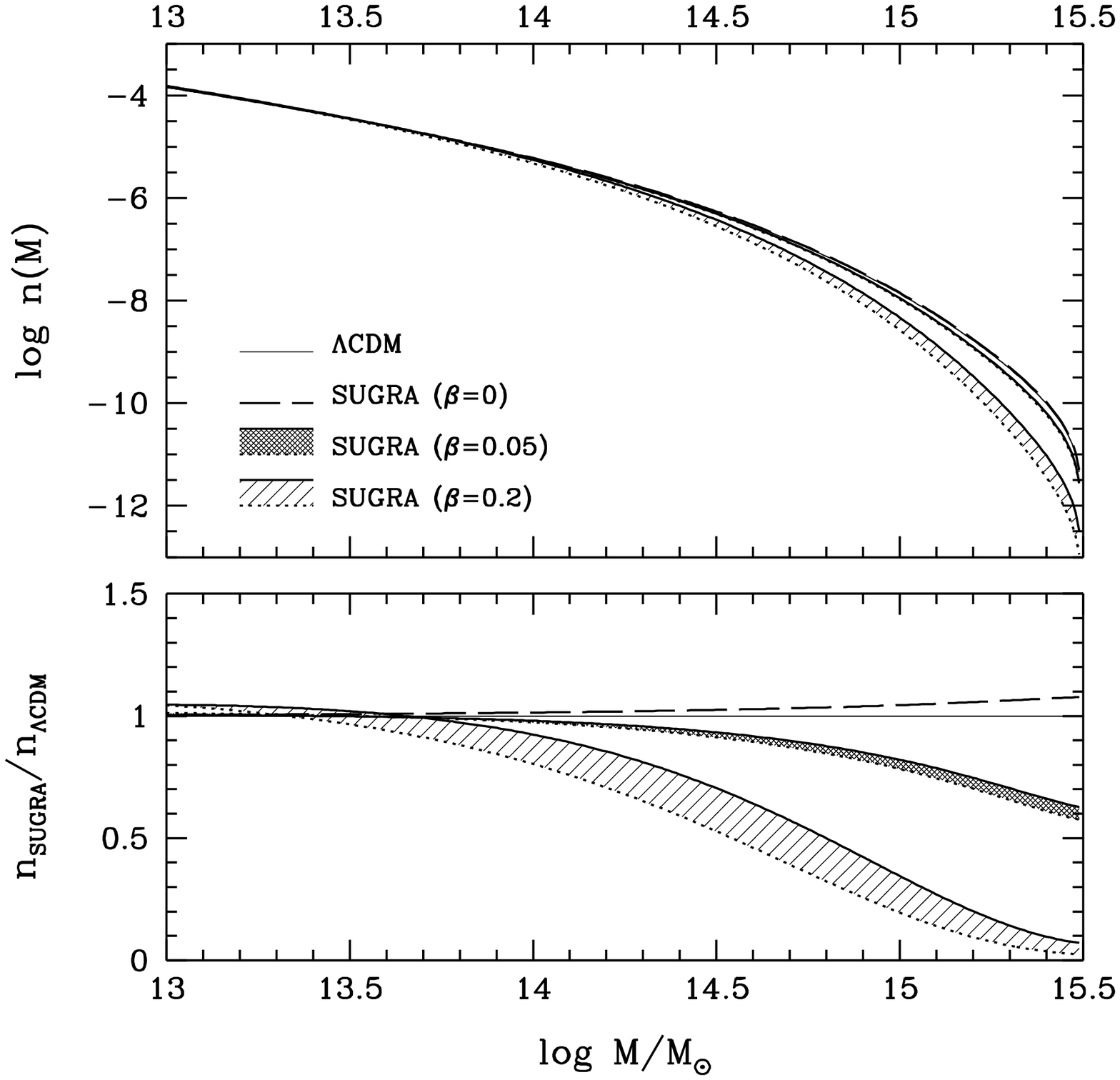}
\includegraphics*[width=6cm]{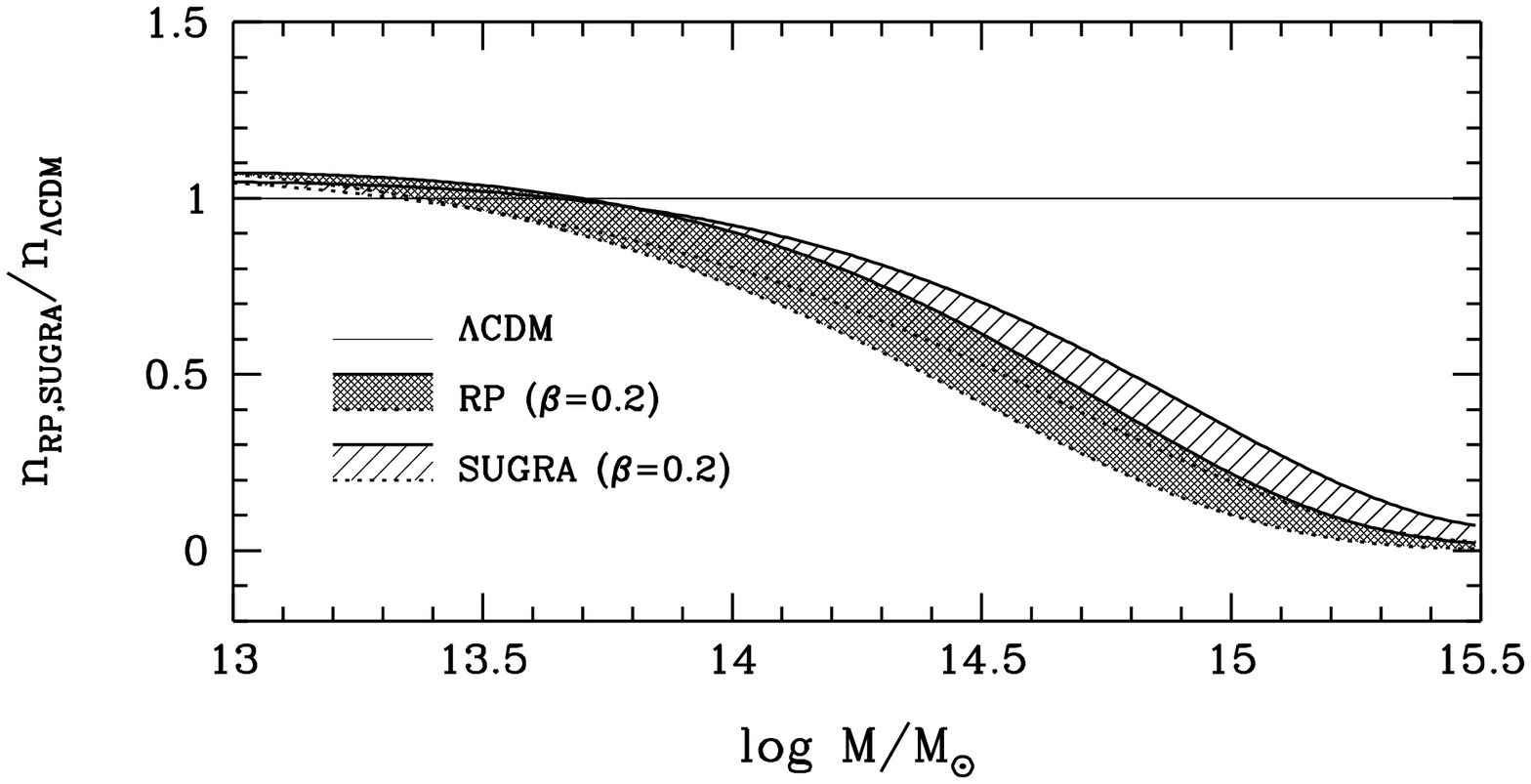}
\caption{Cluster mass function at $z=0$. Four models are
considered: $\Lambda$CDM, SUGRA dDE with $\Lambda = 100\, $MeV, and
two SUGRA cDE with $\beta=0.2$ and 0.05. Dashed areas are limited by
the mass functions for $\delta_{c,rc}^{(c)}$ or $\delta_{c,rc}^{(b)}$
in cDE models. The lower panel yields ratios in respect to $\Lambda$CDM.
The r.h.s. panel overlaps cDE results for SUGRA and RP. }
\label{mfz0}
\end{figure}
We shall plot mass functions obtained using either
$\delta_{c,rc}^{(c)}$ or $\delta_{c,rc}^{(b)}$. Actual data should
fall in the interval between these functions. If, during the process
of cluster formation, outer layers were stripped by close encounters,
data shall be closer to the $\delta_{c,rc}^{(c)}$ curve. 

Figure \ref{mfz0} shows $n(>$$M) = \int_M^\infty dM'\, \psi(M')$
obtained through eq.~(\ref{ps}). The lower panel shows the ratio
between halo numbers for each model and $\Lambda$CDM.  No large
differences between $\Lambda$CDM and dDE are found at $z=0\, $. Shifts
are greater between $\Lambda$CDM and cDE.  For $\beta = 0.20$,
clusters with $M >\sim 3 \cdot 10^{14} h^{-1} M_\odot$ are half than
in $\Lambda$CDM. For $\beta = 0.05$ the shift is smaller, hardly
reaching $20 \, \%$, still opposite to dDE. The r.h.s. panel shows
that these features are true also for RP. More RP results can be found
in \cite{maibon6}.
\begin{figure}
\includegraphics*[width=6cm]{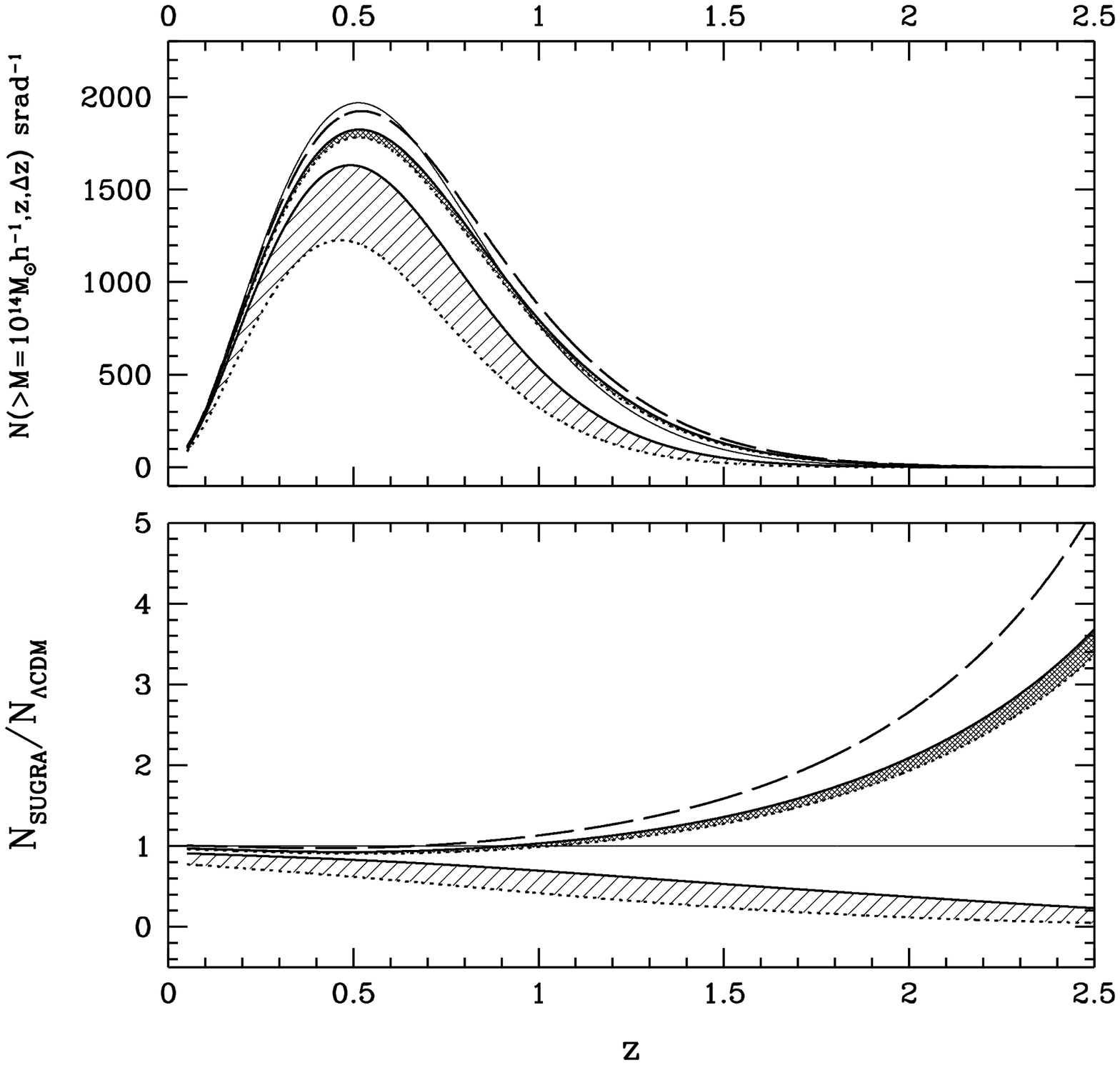}
\includegraphics*[width=6cm]{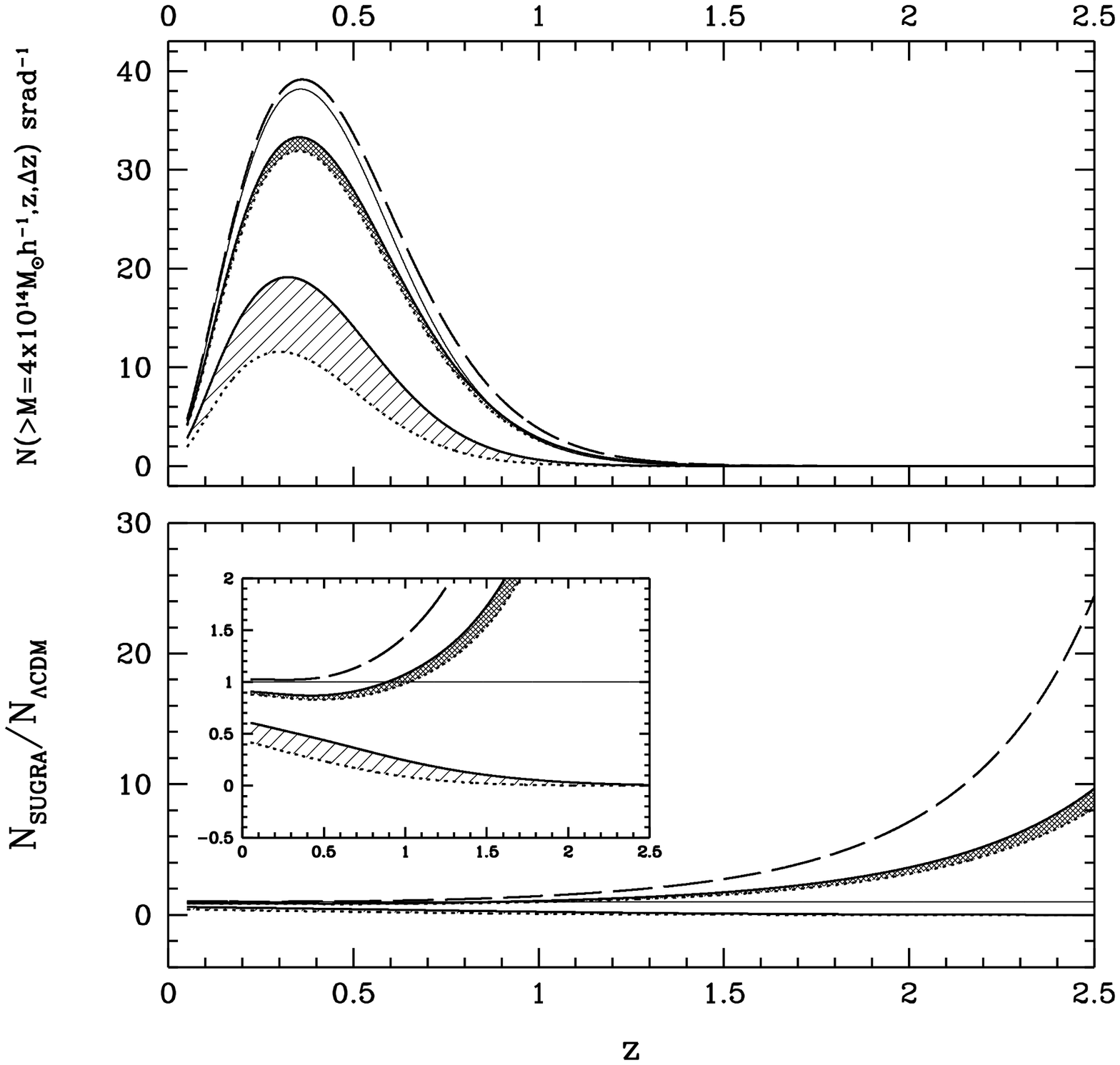}
\caption{Number of clusters with $M>10^{14}h^{-1} M_\odot\, $ (left panel)
or $4 \cdot 10^{14}h^{-1} M_\odot\, $ (right panel) in a fixed solid
angle and redshift interval. The upper linear (lower log) plot
outlines shifts at small (large) $z$.  }
\label{cmf}
\end{figure}

To discriminate between models, cluster evolution predictions should
provide numbers per solid angle and redshift interval (rather than for
comoving volumes, see \cite{sole}) as in the upper panels of
Fig.~\ref{cmf}. In the lower panels, SUGRA models and $\Lambda$CDM are
compared. We consider two masses: $10^{14}$ and $4 \cdot 10^{14}
h^{-1} M_\odot$; for the latter scale, a magnified box shows the
expected low--$z$ behavior showing how we pass from numbers smaller
than $\Lambda$CDM to greater numbers, for $\beta = 0.05$, at a
redshift $z \simeq 0.7\, $. The high--$z$ behaviors for $\beta = 0.05
$ or 0.2 lay on the opposite sides of $\Lambda$CDM.

\vskip .4truecm
\noindent
{\it 5.~~Conclusions}

\vglue .2truecm

\noindent
In this paper we discussed individual cluster formation and cluster
mass functions. A first finding is the expected baryon--DM segregation
causing baryon depletion in clusters. Depletion could be even stronger
if the outer layers are stripped out by close encounters during the
formation process. Preliminary results of simulations confirm these
outputs.

As far as mass functions are concerned, a first significant feature is
that the discrepancy of dDE from $\Lambda$CDM is partially or totally
erased by a fairly small DM--DE coupling, and many cDE predictions lay
on the opposite side of $\Lambda$CDM, in respect to dDE. 

At $z=0$, a shortage of large clusters is expected in cDE. Therefore,
if $\Lambda$CDM is used to fit data in a cDE world, cluster data yield
a $\sigma_8$ smaller than galaxy data.

When we consider the $z$ dependence, we see that (i) when passing from
$\Lambda$CDM to uncoupled SUGRA, the cluster number is expected to be
greater. (ii) When coupling is added, the cluster number excess is
reduced and the $\Lambda$CDM behavior is reapproached. (iii) A
coupling $\beta = 0.05$ may still yield result on the upper side of
$\Lambda$CDM, while $\beta = 0.2$ displaces the expected behavior well
below $\Lambda$CDM.

This can be fairly easily understood. When $\rho_{de}$ keeps constant,
while $\rho_m \propto (1+z)^3$, DE relevance rapidly fades. In dDE
models, instead, $\rho_{de}$ (slightly) increases with $z$ and, to get
the same amount of clusters at $z=0$, they must be there since
earlier. Coupling acts in the opposite way, as gravity is boosted by
the $\phi$ field. In the newtonian language, this means greater
gravity constant ($G^*$) and DM particle masses at high $z$, speeding
up cluster formation: a greater $\beta $ needs less clusters at high
$z$ to meet their present number.

The overall result we wish to outline, however, is that non linearity
apparently boosts the impact of coupling so that there are quite a few
effects from which the coupling between DM and DE can be gauged. Most
of them are just below the present observational threshold and even
slight improvements of data precision will begin to allow
discriminatory measurements. An essential step to fully exploit such
data will be performing n--body and hydro simulations of cDE
models. Through this pattern we can therefore expect soon more
information on the actual nature of the Dark Side.

\end{document}